\documentclass[letterpaper, 10 pt, conference]{ieeeconf}  

\IEEEoverridecommandlockouts                              
\usepackage[latin1]{inputenc}
\usepackage{amsmath}
\usepackage{amsthm}
\usepackage{amsfonts}
\usepackage{amssymb}
\usepackage{graphicx}
\usepackage{cleveref}
\usepackage{caption}
\usepackage{subcaption}
\usepackage{epstopdf}
\usepackage{algorithm}
\usepackage[noend]{algpseudocode}
\usepackage{soul,color}
\usepackage{tikz}
\usetikzlibrary{arrows}
\usetikzlibrary{positioning}

\DeclareMathOperator*{\argmax}{arg\,max}
\theoremstyle{definition}

\newtheorem{proposition}{Proposition}

\newtheorem{remark}{Remark}

\begin{document}
	\title{\LARGE \bf Traffic Automation in Urban Road Networks Using Consensus-based Auction Algorithms For Road Intersections}
	\author{Fabio Molinari, Alexander Martin Dethof, J\"org Raisch
		\thanks{F. Molinari is with the Control Systems Group - Technische Universit\"at Berlin, Germany.}
		\thanks{A. M. Dethof is with the Control Systems Group - Technische Universit\"at Berlin, Germany.}
		\thanks{J. Raisch is with the Control Systems Group - Technische Universit\"at Berlin, Germany \& Max-Planck-Institut f\"ur Dynamik Komplexer Technischer Systeme, Germany.}
		\thanks{\tt\small molinari@control.tu-berlin.de, alexander.m.dethof@campus.tu-berlin.de, raisch@control.tu-berlin.de}
		\thanks{
			This work was funded by the German Research Foundation (DFG) within their priority programme SPP 1914 "Cyber-Physical 
			Networking (CPN)",  RA516/12-1.
		}
	}
	\maketitle
	\begin{abstract}
		This paper 
		describes a decentralized
		control strategy for the automation of 
		road intersections and
		studies its impact
		on traffic in a realistic
		urban road network.
		The controller incorporates
		a consensus-based auction algorithm (CBAA-M),
		which allows vehicles to agree
		on a crossing order
		at each road intersection,
		and an on-board model predictive controller
		that avoids collisions with
		other traffic participants,
		while trying to 
		satisfy performance metrics over time.
		Randomized simulations show 
		that this decentralized control
		approach
		guarantees efficiency, safety, and
		a
		higher throughput than traditional
		solutions.
	\end{abstract}
\section{Introduction}
	In the last decades, 
	the study of fully automated
	vehicular traffic
	has been in the spotlight, 
	see \cite{ioannou1993autonomous, katriniok2013optimal}.
	Cutting the human driver out of the loop
	pledges more safety, 
	higher traffic efficiency,
	and a decrease of air pollution.
	The automation of extra-urban traffic
	has encouraged studies,
	e.g.,
	on
	path following \cite{arogeti2012path} and
	autonomous overtaking with collision avoidance \cite{molinari2017efficient}.
	In an urban environment,
	where overtaking is usually
	not allowed,
	automating
	road intersections represents
	the main challenge.
	Traditionally,
	in the adjacency of an intersection,
	traffic lights
	are in charge of 
	deciding - in a centralized fashion -
	the crossing order of vehicles.
	In the near future,
	when autonomous vehicles will be
	able to exchange information
	(according to the so-called V2V -- vehicle to vehicle -- and V2I -- vehicle to infrastructure -- communication),
	traffic lights can be replaced.
	One possible strategy is to design a 
	centralized optimal controller which gives
	each vehicle a crossing order.
	Appropriate factors, such
	as actual and desired speeds or
	inter-vehicular distances,
	will be weighted while
	trying to maximize the 
	intersections' throughput.
	However,
	employing a centralized
	controller creates real-time
	implementability issues.
	In fact, as problem complexity
	complexity increases exponentially
	with the addition of more intersections
	and more vehicles,
	realistic scenarios defy 
	real-time solutions.
	Moreover,
	in case 
	the centralized controller fails,
	the whole traffic network will be affected.
	
	In \cite{molinari2018automation},
	a possible solution
	employing a decentralized control structure
	for the automation of a road intersection
	is presented.
	It builds
	an algorithm for task assignment from robotics 
	(see \cite{choi2009consensus}).
	In the following, this solution
	will be referred to as
	CBAA-M (Consensus Based Auction Algorithm Modified).
	Vehicles agree on a crossing order
	for each possible collision
	point in the intersection
	by participating in an auction
	without a central auctioneer.
	This is possible by employing
	a consensus protocol
	see e.g. \cite{ren2007information}.
	Inspired by \cite{katriniok2017distributed},
	each vehicle will use an on-board
	Model Predictive Controller (MPC)
	designed to avoid collisions
	with higher priority (and frontal) vehicles
	while minimizing some cost.
	This fully decentralized control strategy
	was shown to guarantee a real-time collision-free
	solution to the problem
	while providing a high throughput.
	
	The following work aims to extend the results
	presented in \cite{molinari2018automation}
	to the case of urban road networks with many intersections.
	In Section \ref{sec:probDesc},
	a realistic urban scenario, composed of 
	a collection of adjacent road intersections
	is described. 
	In Section \ref{sec:CBAAM},
	CBAA-M is reviewed and proven to achieve
	finite-time convergence
	for groups of cars with a connected
	(but not necessarily fully connected)
	network topology.
	The optimal control structure is introduced in Section \ref{Sec:MPC};
	differently from \cite{molinari2018automation},
	vehicles are also allowed to turn left
	at intersections.
	Section \ref{sec:simulations}
	presents simulation results
	and quantifies the impact
	of this solution.
	Finally, an analysis of the impact
	of disallowing left turning
	is presented.
	
	\subsection{Notation}
	\begin{figure}[t]
		\includegraphics[width=\columnwidth]{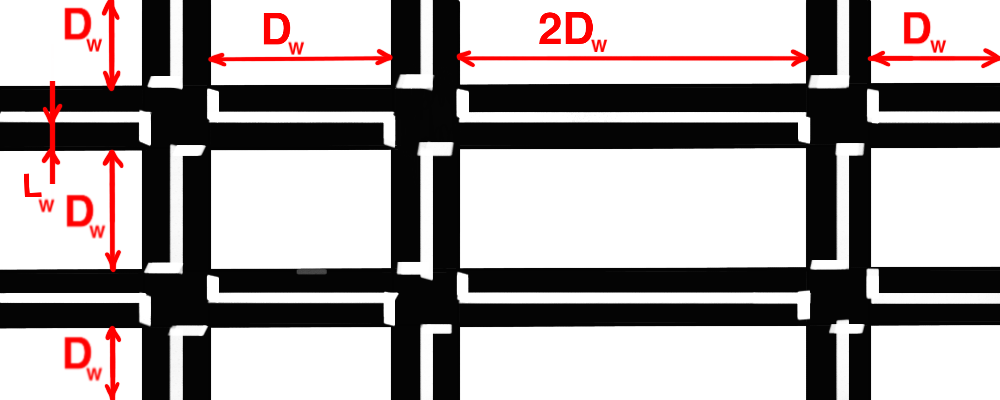}
		\caption{Urban road network. $L_W$ is the width of the lane, while 
			the length of each road sector is an integer multiple of $D_W$.}
		\label{fig:roadNet}
	\end{figure}
	Throughout the paper,
	$\mathbb{N}_0$ denotes the set of nonnegative integers,
	$\mathbb{R}$ the set of real numbers,
	and $\mathbb{N}$ the set of positive integers.
	The set of nonnegative and positive real numbers are,
	respectively,
	$\mathbb{R}_{\geq0}$ and $\mathbb{R}_{>0}$.
	Given a set $\mathcal{S}$, its cardinality is $|\mathcal{S}|$.
	An undirected graph
	is a pair $(\mathcal{S},\mathcal{A})$,
	where $\mathcal{S}$ is a set of nodes
	and $\mathcal{A}\subseteq[\mathcal{S}]^2$ a set of arcs,
	with $[\mathcal{S}]^2$ denoting the set of all two-element
	subsets of $S$.
	Given a node $i\in\mathcal{S}$,
	its neighbors set is 
	$\mathcal{S}_i=\{ 
		j \mid \{i,j\}\in\mathcal{A}
	 \}$.
	A path between nodes $i_1\in\mathcal{S}$ and $i_n\in\mathcal{S}$
	$i_1\not=i_n$,
	is a set of arcs 
	of the form $\amalg=\{ \{i_1,i_2\},\{i_2,i_3\},\dots \{i_{n-2},i_{n-1}\},\{i_{n-1},i_n\} \}$.
	The graph $(\mathcal{S},\mathcal{A})$ 
	is connected if
	there exists a path 
	between any pair of nodes $i,j$, $i\not=j$.
	It is fully connected,
	if there is an arc between any two 
	nodes, i.e., $\mathcal{A}=[\mathcal{S}]^2$.
	The i-th entry
	of vector $\mathbf{v}$
	is $(\mathbf{v})_i$.
	Two sorting functions are used.
	The function $\text{sort} : \mathbb{R}^n_{\geq0}\rightarrow\mathbb{R}^n_{\geq0}$
	organizes vector elements in decreasing order
	of magnitude; the function $\text{argsort} : \mathbb{R}^n_{\geq0}\rightarrow\mathbb{N}^n$ 
	displays vector indices in decreasing order of their respective entries' magnitude. 
	The function
	$\text{argmax} : \mathbb{R}^n_{\geq0}\rightarrow\mathbb{N}$
	yields the index 
	of the maximal entry
	in the vector. 
	If 
	more than one entry
	has maximum value, 
	one index is selected 
	randomly among the possible
	candidates.
\section{Problem Description}
\label{sec:probDesc}
An \textit{urban road network} 
is a system of interconnected
roads 
which are designed to carry vehicular traffic,
as in Figure \ref{fig:roadNet}.
Each road is composed of two lanes,
one per direction.
For simplicity and without loss of generality,
only a \textit{Manhattan-like}
grid,
in which all roads are perpendicularly intersecting,
will be analyzed.
Overtaking and \textit{u-turns} are forbidden.
The traffic is composed of fully autonomous vehicles
communicating with each other (V2V).
The availability
of a common shared clock is assumed.
	\subsection{Vehicle model}
	\label{sec:vehModel}
		A set $\mathbf{N}=\{1,\dots,n \}$ of $n>1$ vehicles driving on
		the urban road network is considered.
		Each vehicle, say $i\in\mathbf{N}$, is given
		a desired path in global coordinates,
		i.e. $\mathcal{P}_i=\{ \mathbb{X}_i^g, \mathbb{Y}_i^g \}\subset\mathbb{R}^2$,
		which comes from a higher-level GPS navigation system.
		The latter provides vehicle
		$i$ also with a desired cruising speed, 
		i.e. $v_i^{r}\in\mathbb{R}_{>0}$.
		Each vehicle $i\in\mathbf{N}$ is modeled,
		as in \cite{murgovski2015convex},
		by a point-mass 
		discrete-time 
		linear 
		system. 
		Let $\mathbf{x}_i(k)=[p_i(k),v_i(k)]'$ be its state vector,
		where $p_i(k)\in\mathbb{R}_{>0}$ and 
		$v_i(k)\in\mathbb{R}_{\geq0}$
		are,
		respectively,
		position 
		and velocity
		along $\mathcal{P}_i$
		at discrete time instant $k\in\mathbb{N}_0$;
		\begin{equation}
			\label{eq:pointMassLinSys}
			\mathbf{x}_i(k+1)=A\mathbf{x}_i(k) + Bu_i(k),
		\end{equation}
		where
		\begin{equation}
			A = 
			\begin{bmatrix}
				1 &T_s\\
				0 &1
			\end{bmatrix}
			,\ 			
			B = 
			\begin{bmatrix}
				0 \\
				T_s
			\end{bmatrix},
		\end{equation}
		$T_s\in\mathbb{R}_{>0}$ is the sampling time,
		$\mathbf{x}_i(0)=\mathbf{x}_{i_0}$,
		and $u_i:\mathbb{N}\rightarrow\mathbb{R}$ is the 
		longitudinal
		acceleration of vehicle $i$ along $\mathcal{P}_i$.		
		In the following, 
		we refer to $p_i$ as position in \textit{local coordinates}.
		Each vehicle $i$ will have
		a local to global map
		$\mathcal{M}_i:\mathbb{R}_{\geq0}\rightarrow\mathcal{P}_i$,
		which will associate $p_i$ to the respective position
		in the global frame, i.e.
		\begin{equation}
			\label{eq:mapLocToGlob}
			\mathcal{M}_i(p_i(k)) = \mathbf{x}_i^g(k) = 
			(x_i^g(k),\ y_i^g(k))\in\mathcal{P}_i.
		\end{equation}
		The global to local map is also defined and denoted by 
		$\mathcal{M}_i^{-1}:\mathcal{P}_i\rightarrow\mathbb{R}_{\geq0}$.		
		Given two vehicles $i,j$ and a discrete-time index $k\in\mathbb{N}_0$,
		their distance is defined as
		\begin{multline}
			\label{eq:distanceVeh}
			d_{i,j}^k=
			d(\mathbf{x}_i^g(k),\ \mathbf{x}_j^g(k))=\\
				\sqrt{
						(x_i^g(k)-x_j^g(k))^2 + (y_i^g(k)-y_j^g(k))^2
					}.
		\end{multline}
		We say that, at time $k$, 
		a collision between vehicle $i$ and $j$ occurs
		if $d_{i,j}^k<\underline{d}$,
		where $\underline{d}\in\mathbb{R}_{>0}$
		is a given minimum allowed distance.
	\subsection{Frontal Vehicles}
		Clearly, different vehicles can simultaneously 
		drive along the same lane.
		It is important that each vehicle $i$ recognizes
		its current frontal vehicles,
		so that any possible bumper-to-bumper collision
		can be avoided.
		Let $\mathcal{F}_i^k$ be the set of $i$'s frontal vehicles
		at time $k$,
		formally
		\begin{multline}
			\label{eq:frontalSet}
			\mathcal{F}_i^k = 
			\{
				j\in\mathbf{N} 
				\mid
				\mathcal{M}_j(p_j(k))\in\mathcal{P}_i,\\
				\mathcal{M}_i^{-1}( \mathcal{M}_j(p_j(k)) ) > p_i(k)
			\}.
		\end{multline}
	
	\subsection{Collision Points}
	\label{subsec:collPoints}
	
		\begin{figure}[t]
			\vspace{10px}
			\includegraphics[width=\columnwidth]{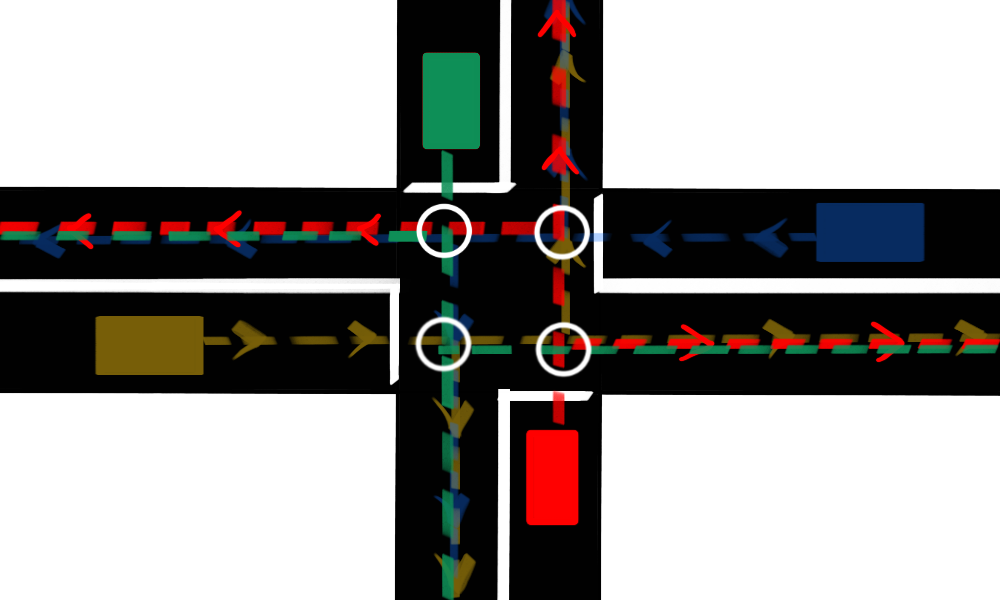}
			\caption{Possible collision points in one intersection.}
			\label{fig:collPoints}
		\end{figure}
		Vehicle $i$ should be aware
		of all those vehicles driving on different road sections
		but going to intersect $\mathcal{P}_i$.
		All the possible paths, i.e. $\{\mathcal{P}_i \mid i\in\mathbf{N} \} $, will determine
		the set of all possible
		\textit{collision points} in the network (cf. Figure \ref{fig:collPoints}), 
		\begin{equation}
			\label{eq:collPointSet}
			\mathcal{H} = 
			\{	
				(h_{x},\ h_{y}) \in \mathbb{R}^2
			\}.
		\end{equation}
		Let $G_i^k$ be the set of all collision points that vehicle $i$
		has still to cross at time $k$, i.e.
		\begin{equation}
			\label{eq:G_i^k}
			G_i^k
			=
			\{
				h \in \mathcal{H}
				\mid
				h \in \mathcal{P}_i,\ 
				\mathcal{M}_i^{-1}(h) > p_i(k)
			\}.
		\end{equation}
		On the other hand,
		given $h\in\mathcal{H}$,
		the set $H_{h}^k$ collects
		all the vehicles that still have to cross,
		at time $k$, the collision point $h$, i.e.,
		\begin{equation}
			\label{eq:Hrjk}
			H_{h}^k = 
			\{
				i\in\mathbf{N}
				\mid
				h\in\mathcal{H},\
				h\in\mathcal{P}_i,\
				\mathcal{M}_i^{-1}(h) > p_i(k)
			\}.
		\end{equation}
		Let $\mathcal{L}_i^k$ be the set
		of all vehicles with 
		a higher priority than vehicle $i$
		for crossing the collision point in $G_i^k$.
		A method to compute $\mathcal{L}_i^k$ 
		in a complete decentralized fashion
		will be
		presented in Section \ref{sec:CBAAM}.
	\subsection{Control Structure}
	\label{sec:controlStruct}
	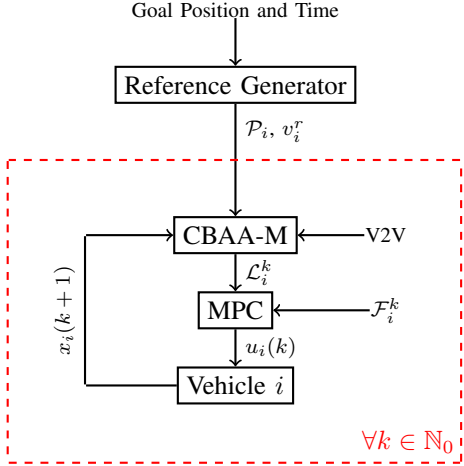
\begin{figure}[h]
		\vspace{10px}
		\centering
		\begin{tikzpicture}[thick]		
			\node[inner sep=0,minimum size=0] (l01) {\footnotesize Goal Position and Time}; 
			
			\node[draw,rectangle, below of=l01] (a) {Reference Generator};
			
			\node[inner sep=0,minimum size=0,below of=a] (b01) {}; 
			
			\node[draw,rectangle,below of=b01] (b) {CBAA-M};
			\node[draw,rectangle,below of=b] (c) {MPC};
			\node[draw,rectangle,below of=c] (d) {Vehicle $i$};
			
			\node[inner sep=0,minimum size=0,left of=b] (k01) {}; 
			\node[inner sep=0,minimum size=0,left of=d] (k02) {}; 
			\node[inner sep=0,minimum size=0,left of=k01] (k1) {}; 
			\node[inner sep=0,minimum size=0,left of=k02] (k2) {}; 
			
			\node[inner sep=0,minimum size=0,right of=b] (h01) {}; 
			\node[inner sep=0,minimum size=0,right of=h01] (h02) {\footnotesize V2V}; 
			
			\node[inner sep=0,minimum size=0,right of=c] (m01) {}; 
			\node[inner sep=0,minimum size=0,right of=m01] (m02) {\footnotesize $\mathcal{F}_i^k$}; 
			
			\node[inner sep=0,minimum size=0,left=3cm of b01] (dash1) {}; 
			\node[inner sep=0,minimum size=0,below=4cm of dash1] (dash2) {}; 
			\node[inner sep=0,minimum size=0,right=6cm of dash2] (dash3) {}; 
			\node[inner sep=0,minimum size=0,above=4cm of dash3] (dash4) {}; 
			
			\draw[-] (a) -- (b01) node[right, midway] {\footnotesize $\mathcal{P}_i$, $v_i^r$};
			\draw[->] (b01) -- (b);
			\draw[->] (b) -- (c) node[right,midway] {\footnotesize $\mathcal{L}_i^k$};
			\draw[->] (c) -- (d) node[right,midway] {\footnotesize $u_i(k)$};
			\draw[-] (d) -- (k2);
			\draw[-] (k2) -- (k1) node[above, rotate=90, midway] {\footnotesize $x_i(k+1)$};
			\draw[->] (k1) to (b);			
			\draw[->] (h02) to (b);			
			\draw[->] (l01) to (a);	
			\draw[->] (m02) to (c);
			
			\draw[dashed, red] (dash1) -- (dash2);
			\draw[dashed, red] (dash2) -- (dash3) node[above, xshift=-0.7cm] {$\forall k\in\mathbb{N}_0$};
			\draw[dashed, red] (dash3) -- (dash4);
			\draw[dashed, red] (dash4) -- (dash1);
			
		\end{tikzpicture}
		\caption{Hierarchical control structure for vehicle $i$.}
		\label{fig:hier_ctrl}
	\end{figure}
		This information plays a rucial role
		within the hierarchically structured controller
		for vehicle $i$ (see Figure \ref{fig:hier_ctrl}).
		The ingredients of this controller
		are described below:
		\paragraph{Reference Generator}
			Its task is providing the on-board controller with 
			the desired path and cruising speed,
			given the goal position.
			It is outside the scope
			of this paper.
		\paragraph{CBAA-M}
			this layer runs the so-called \textit{Consensus-based Auction Algorithm Modified}
			(see Section \ref{sec:CBAAM}). 
			Vehicle $i$ will run this algorithm at every discrete-time step $k$
			thus 
			providing the controller with the set $\mathcal{L}_i^k$
			of higher priority vehicles, 
			introduced in Section \ref{subsec:collPoints}.
		\paragraph{MPC}
			this on-board Model Predictive Controller
			minimizes a cost function (reflecting deviations from the desired speed, discomfort, etc)
			while avoiding collisions with vehicles in set $\mathcal{F}_i^k\cup\mathcal{L}_i^k$.			
			Its output is the acceleration $u_i(k)$. 
			
			The proposed structure builds on
			the idea of avoiding collisions only with 
			higher priority and frontal vehicles,
			which was originally
			proposed in \cite{katriniok2017distributed}.
%
		
\section{Consensus-based Auction Algorithm Modified}
	\label{sec:CBAAM}
	CBAA-M is an algorithm that allows a 
	multi-agent system 
	modeled by a graph
	$(\mathcal{S},\mathcal{A})$ 
	to
	achieve an agreement 
	between a set of agents.
	of agents.
	It has been presented in \cite{molinari2018automation},
	which, in turn, takes inspiration from \cite{choi2009consensus}.
	This algorithm is able to run an auction
	without requiring the presence of any central auctioneer.
	It can therefore be employed in a completely decentralized control approach,
	as the one adopted in our strategy.		
	In the following, 
	we will briefly summarize CBAA-M
	and prove that
	it converges
	for connected graphs. 
	This is a sharper result than in
	\cite{molinari2018automation},
	where full connectedness was
	assumed.
	
	Each agent $i\in\mathcal{S}=\{1,\dots,S\}$
	places a bid $c_i\in\mathbb{R}_{>0}$
	that determines
	its position in the sequence. 
	We assume that each agent has a distinct bid to place, 
	i.e. $\forall i\not=j,\ c_i\not=c_j$.
	The higher the bid, 
	the earlier that agent could appear in the resulting sequence.
	Each agent has two vectors of dimension $S$, i.e. $\mathbf{v}_i$ and $\mathbf{w}_i$,
	which are updated at each iteration $\kappa\in\mathbb{N}$
	and are initialized as zero vectors.
	The first vector (winners list)
	has to be filled with a sequence of agents' indexes,
	whilst the second vector
	contains their respective bids.
	This algorithm is composed of two subsequent 
	phases within every iteration step $k$:
	\begin{itemize}
		\item[1)] Local Auction [\textit{Algorithm \ref{cbaam1}}]: at each iteration $\kappa$, 
		each agent $i\in\mathcal{S}$ places, if its index is not already stored in $\mathbf{v}_i^\kappa$,
		its own bid $c_i$ in the earliest possible position
		of vector $\mathbf{w}_i^\kappa$.
		In the same position, it stores its index in vector $\mathbf{v}_i^\kappa$.
	\end{itemize}	
	\begin{algorithm}
		\caption{CBAA-M Phase 1 for agent $i$ at iteration $\kappa$}\label{cbaam1}
		\label{algo:cbaam1}
		\begin{algorithmic}[1]
			\State $\mathbf{v}_i^{0}=\mathbf{0}_S$ , $\mathbf{w}_i^{0}=\mathbf{0}_S$
			\Procedure{Bid($c_i$,\ $\mathbf{v}_i^{\kappa-1}$,\ $\mathbf{w}_i^{\kappa-1}$)}{}
			\State $\mathbf{v}_i^{\kappa}\gets \mathbf{v}_i^{\kappa-1}$
			\State $\mathbf{w}_i^{\kappa}\gets \mathbf{w}_i^{\kappa-1}$
			\State $j\gets 1$
			\State \emph{loop}:
			\If {$i\not=(\mathbf{v}_{i}^{\kappa})_l$, $\forall l=1\dots S$}
			\If {$c_i>(\mathbf{w}_{i}^{\kappa-1})_j$}
			\State $(\mathbf{v}_{i}^{\kappa})_j \gets i$
			\State $(\mathbf{w}_{i}^{\kappa})_j \gets c_i$
			\EndIf
			\State $j\gets j+1$		
			\State \textbf{goto} \emph{loop}.
			\EndIf
			\State \textbf{close};
			\EndProcedure
		\end{algorithmic}
	\end{algorithm}		
	\begin{itemize}
		\item[2)] Consensus over the lists [\textit{Algorithm \ref{cbaam2}}]: after the first phase,
		each agent has its own version of $\mathbf{v}_i^\kappa$ and $\mathbf{w}_i^\kappa$.
		The network need to reach an agreement on them.
		For that purpose, each agent $i\in\mathcal{S}$
		sends its vectors to its respective neighbors 
		(i.e. the nodes in the set $\mathcal{S}_i$)
		and receives theirs.
		Then, via a max-consensus protocol,
		it selects the best bid for each row of $\mathbf{w}_i^\kappa$
		and puts in the same position of $\mathbf{v}_i^\kappa$ the respective
		agent's index.
	\end{itemize}
		
	\begin{algorithm}
		\caption{CBAA-M Phase 2 for agent $i$ at iteration $\kappa$}\label{cbaam2}
		\begin{algorithmic}[1]	
			\label{algo:cbaam2}		
			\State $\text{SEND}$ $(\mathbf{v}_i^{\kappa},\mathbf{w}_i^{\kappa})$ to $j\in{\mathcal{S}}_i=\{j\in\mathcal{S}\mid (i,j)\in\mathcal{E}\}$
			\State $\text{RECEIVE}$ $(\mathbf{v}_h^{\kappa},\mathbf{w}_h^{\kappa})$ from $h\in{{\mathcal{S}}}_i=\{j\in\mathcal{S}\mid (j,i)\in\mathcal{E}\}$
			\Procedure{Update($\mathbf{v}_{h\in{\mathcal{S}}_i}^{\kappa}$, $\mathbf{w}_{h\in{\mathcal{S}}_i}^{\kappa}$)}{}
			\State $(\mathbf{{a}}_{i}^{\kappa})_j \gets \argmax\limits_{h\in\mathcal{S}_i}((\mathbf{w}_{h}^{\kappa})_j),\ \forall j=1\dots \max(\kappa,S)$
			\State $(\mathbf{v}_{i}^{\kappa})_j \gets (\mathbf{v}_{(\mathbf{{a}}_{i}^{\kappa})_j}^{\kappa})_j\quad\forall j=1\dots \max(\kappa,S)$
			\State $(\mathbf{w}_{i}^{\kappa})_j \gets \max\limits_{h\in\mathcal{S}_i}((\mathbf{w}_{h}^{\kappa})_j)\quad \forall j=1\dots \max(\kappa,S)$
			\EndProcedure
		\end{algorithmic}
	\end{algorithm}	
	After terminating Phase 2,
	$(\mathbf{w}_{h}^{\kappa})_j$
	is the maximal bid for
	position $j$
	that agent $i$ 
	is aware of,
	and $(\mathbf{v}_{h}^{\kappa})_j$
	is the index of the agent
	having placed that bid.
	
	In \cite{molinari2018automation}, 
	a fully connected network topology was assumed and
	an agreement was 
	shown to be
	reached in exactly $S$ iterations.
	In the following,
	we will show that
	the algorithm converges 
	under the 
	weaker condition
	of $(\mathcal{S},\mathcal{A})$
	being
	connected.
	\begin{proposition}
		A multi-agent system 
		represented by a connected graph
		$(\mathcal{S},\mathcal{A})$
		executes CBAA-M.
		An agreement is reached in $\bar{\kappa}\in\mathbb{N}$ iterations.
		Formally, $\exists \bar{\kappa}\in\mathbb{N}$: 
		\begin{align}
			\label{eq:propagationMax_proposition1}
			\forall i,j\in\mathbb{N},\ \ \  &
			\mathbf{v}_{i}^{\bar{\kappa}}=\mathbf{v}_{j}^{\bar{\kappa}}=\mathbf{v}^*=\text{argsort}(\mathbf{c}),\\			
			\label{eq:propagationMax_proposition2}
			& \mathbf{w}_{i}^{\bar{\kappa}}=\mathbf{w}_{j}^{\bar{\kappa}}=\mathbf{w}^*=\text{sort}(\mathbf{c}),
		\end{align}
		where, $\forall i\in\mathcal{S},\ (\mathbf{c})_i=c_i$.
		
		\begin{proof}
			Given a pair $i\in\mathcal{S},\ \kappa\in\mathbb{N}$,
			such that $\mathbf{v}_i^\kappa=\mathbf{v}^*$ and $\mathbf{w}_i^\kappa=\mathbf{w}^*$,
			the following holds:
			\begin{equation}
				\label{eq:propagationMax}
				\forall j\in\mathcal{S}_i\cup\{i\},\ 
				\mathbf{v}_j^{\kappa+1}=\mathbf{v}^*,\ 
				\mathbf{w}_j^{\kappa+1}=\mathbf{w}^*.
			\end{equation}
			In fact, 
			if (\ref{eq:propagationMax}) does not hold,
			Algorithm \ref{cbaam2}
			would imply that, for an arbitrary $j\in\mathcal{S}_i$, 
			$\exists g=1,\dots,S,\ (\mathbf{w}_j^{\kappa})_g>(\mathbf{w}^*)_g$.
			The latter, by Algorithm \ref{cbaam1}, 
			means that $\exists h=1,\dots,S:\ c_g<{c}_h<c_{g+1}$,
			which is in contradiction with the definition of $\mathbf{w}^*$,
			i.e. $\mathbf{w}^*=\text{sort}(\mathbf{c})$.			
			A similar analysis can be conducted for $\mathbf{v}_j^\kappa$.
			
			The agreement phase of CBAA-M follows
			a max-consensus protocol for each row of $\mathbf{v}_i$ and $\mathbf{w}_i$.
			By \cite{nejad2009max}, 
			max-consensus is achieved in a connected network
			in at most $l$ steps,
			where
			\begin{equation}
				\label{eq:stepsToMax}
				l=\max_{i,j\in\mathcal{S}}\{ |i,j|_{l,min} \},
			\end{equation}
			where $|i, j|_{l,min}$ denotes the length of the shortest existing
			path between node $i$ and node $j$.
			Therefore, by (\ref{eq:propagationMax}) and (\ref{eq:stepsToMax}),
			if agent $i_0$ at iteration $\kappa_0$ is the first agent to have
			$\mathbf{v}_{i_0}^{\kappa_0}=\mathbf{v}^*$ and $\mathbf{w}_{i_0}^{\kappa_0}=\mathbf{w}^*$,
			then $\bar{\kappa}\leq \kappa_0 +l$.
						
			Let's now estimate $\kappa_0$ in the worst case scenario.
			By Algorithm \ref{cbaam1}, 
			intuitively,
			${i}_0$ is the first agent
			to fill the last row of $\mathbf{w}_i$.
			This is only possible, by Algorithm \ref{cbaam1}, 
			if $i_0$ is not contained in $\mathbf{v}_{i}$
			and 
			if the first $S-1$ rows of $\mathbf{w}_i$ have already been filled.			
			By this, agent $i_0$ must be
			the agent having the minimum bid in the network,
			i.e.
			$c_{{i}_0}=\min_{i\in\mathcal{S}}c_i$.
			In the worst case scenario, 
			by \cite{nejad2009max} and (\ref{eq:stepsToMax}),
			for retrieving the information of each row $j$ 
			of $\mathbf{v}_i$ and $\mathbf{w}_i$,
			agent $i$ needs
			maximum $l$ steps.
			Therefore, always in the worst case scenario,
			$\kappa_0\leq (S-1)l$.
			Finally,
			we can state that 
			the multi-agent system achieves
			an agreement as (\ref{eq:propagationMax_proposition1})-(\ref{eq:propagationMax_proposition2})
			in $\bar{\kappa}$ steps, where
			$$\bar{\kappa}\leq \kappa_0+l \leq (S-1)l+l=Sl.$$
			This concludes the proof.
			\qed
		\end{proof}
	\end{proposition}
	Due to the need of real-time implementation, 
	a fast convergence to the agreement vector
	is often required.
	In \cite{molinari2018exploitingmax}, 
	it was shown
	how
	the convergence rate
	of a max-consensus protocol
	can be drastically 
	increased by 
	harnessing the interference

As claimed in Section \ref{sec:controlStruct},
each vehicle $i\in\mathbf{N}$, 
at every discrete time step $k\in\mathbb{N}_0$,
runs one CBAA-M for each collision point $h\in G_i^k$,
thus retrieving
$|G_i^k|$ priority vectors.
These can be grouped in a set 
$
\{
	\mathbf{v}_h^* \mid h\in G_i^k
\}
$,
where $\mathbf{v}_h^*\in\mathbb{N}^{|H_h^k|}$
is the agreed crossing priority list for collision point $h$.
From this collection of vectors,
vehicle $i$ can retrieve the set $\mathcal{L}_i^k$ (cf. Section \ref{subsec:collPoints})
that collects all the vehicles having higher priority
than $i$ 
at some collision points.
Formally, 
\begin{equation}
	\label{eq:priorityList}
	\mathcal{L}_i^k
	=
	\{
	(\mathbf{v}_h^*)_j 
	\mid 
	h\in G_i^k,\
	j<m,\
	(\mathbf{v}_h^*)_m=i
	\}.
\end{equation}
As in \cite{molinari2018automation}, 
the bid $c_{i,h}^k$ placed by vehicle $i$ at time $k$
in the auction for crossing $h\in G_i^k$
is determined from
its current velocity
and its distance from $h$:
\begin{equation}
	\label{eq:bidDynamics}
	c_{i,h}^k 
	=
	\frac{
		p_v v_i(k)+p_d
		}{d_{i,h}^k+\epsilon},
\end{equation}
where $p_v\in\mathbb{R}_{>0}$, $p_d\in\mathbb{R}_{>0}$, and
$\epsilon\in\mathbb{R}_{>0}$ are chosen parameters
(equal for all vehicles).
The quantity $d_{i,h}^k$ is the distance at time $k$
from vehicle $i$ to the collision point $h\in\mathcal{H}$, 
i.e.,
$d_{i,h}^k=
\sqrt{
	(x_i^g(k)-h_x)^2 + (y_i^g(k)-h_y)^2
}$.
The reader can refer to \cite{molinari2018automation}
for a comprehensive analysis of
$c_{i,h}^k $ and potential coherency problems
in the auction procedures.

\section{Onboard Model Predictive Controller}
\label{Sec:MPC}
As shown in Figure \ref{fig:hier_ctrl}, 
each vehicle $i\in\mathbf{N}$ has an on-board MPC controller
that computes an optimal longitudinal acceleration 
at every discrete step $k\in\mathbb{N}_0$.
It is provided with the desired path $\mathcal{P}_i$,
the desired cruising speed $v_i^r$,
the current vehicle state $\mathbf{x}_i(k)$,
the sets $\mathcal{L}_i^k$ and $\mathcal{F}_i^k$,
and
the states $\mathbf{x}_j(k)$ of all vehicles $j\in\mathcal{L}_i^k\cup\mathcal{F}_i^k$.
	\subsection{Prediction Model}
		MPC 
		employs a prediction model
		of the traffic situation
		extrapolated from the current
		states
		and
		evolving
		along a horizon $T_h\in\mathbb{N}$.
		In the following,
		we describe
		the optimal control problem to be
		solved at every $k\in\mathbb{N}_0$
		by vehicle $i\in\mathbf{N}$.
		
		The predicted state and input
		of vehicle $i$ itself
		are 
		$\tilde{\mathbf{x}}_i(t)=[\tilde{p}_i(t),\tilde{v}_i(t)]$
		and $\tilde{u}_i(t),\ t\in[0,T_h]$.
		The prediction model evolves 
		according to
		\begin{equation}
			\label{eq:predModel_i}
			\tilde{\mathbf{x}}_i(t+1)=A\tilde{\mathbf{x}}_i(t)+B\tilde{u}_i(t),
		\end{equation}
		where $t\in[0,T_h-1]$.
		The prediction model
		of other vehicles
		is based on a constant acceleration assumption:
		in fact,
		$\forall \zeta\in\mathcal{F}_i^k\cup\mathcal{L}_i^k$,		
		\begin{equation}
			\label{eq:predModel_zeta}
			\tilde{\mathbf{x}}_\zeta(t+1)=A\tilde{\mathbf{x}}_\zeta(t) + B\tilde{u}_\zeta,
		\end{equation}
		where $t\in[0,T_h-1]$ and $\tilde{u}_\zeta=u_\zeta(k)$.
		Clearly, 
		the prediction variables are initialized
		according to the current measurement
		at instant $k$, i.e.,
		\begin{align}
			\label{eq:initialCond1}
			&\tilde{\mathbf{x}}_i(0)=\mathbf{x}_i(k),\\
			\label{eq:initialCond2}
			\forall\zeta\in\mathcal{L}_i^k\cup\mathcal{F}_i^k,\ &\tilde{\mathbf{x}}_\zeta(0)=\mathbf{x}_\zeta(k).
		\end{align}
		\begin{remark}
			Let $\mathcal{F}_i^t,\ \forall t\in[0,T_h]$, be
			the set of vehicles $j\in\mathcal{L}_i^k\cup\mathcal{F}_i^k$
			that are in front of vehicle $i$ at prediction time $t$.
			Formally, as in (\ref{eq:frontalSet}),			
			\begin{multline}
				\label{eq:frontalSetPred}
				\mathcal{F}_i^t = 
				\{
				j\in\mathcal{L}_i^k\cup\mathcal{F}_i^k
				\mid
				\mathcal{M}_j(\tilde{p}_j(t))\in\mathcal{P}_i,\\
				\mathcal{M}_i^{-1}( \mathcal{M}_j(\tilde{p}_j(t)) ) \geq \tilde{p}_i(t)
				\}.
			\end{multline}
			Since any vehicle 
			in $\mathcal{L}_i^k\cup\mathcal{F}_i^k$
			always has priority
			in front of $i$,
			requiring $\mathcal{M}_i^{-1}( \mathcal{M}_j(\tilde{p}_j(t)) ) \geq \tilde{p}_i(t)$ 
			in (\ref{eq:frontalSetPred})
			is redundant.
			Accordingly, (\ref{eq:frontalSetPred}) can be rewritten as
			\begin{equation}
				\label{eq:frontalSetPredRewritten}
				\mathcal{F}_i^t = 
				\{
				j\in\mathcal{L}_i^k\cup\mathcal{F}_i^k
				\mid
				\mathcal{M}_j(\tilde{p}_j(t))\in\mathcal{P}_i
				\},
			\end{equation}
			thus making it clear that 
			the usage of $\mathcal{F}_i^t$ 
			does not affect the convexity
			of the problem, 
			since it does not depend on the choice of $\tilde{u}_i(t)$.
		\end{remark}
	\subsection{Obstacle Avoidance}
		We employ,
		similarly to 
		\cite{katriniok2017distributed},
		the idea of
		avoiding collisions
		with
		higher priority and frontal vehicles.		
		The safety distance to be kept
		is defined via the \textit{continuous-time headway rule},
		as in \cite{chien1992automatic}.
		Then, a general
		collision avoidance constraint
		between vehicle
		$i$ and an arbitrary
		vehicle $\zeta$ 
		at prediction time $t$
		is of the form
		\begin{equation}
			\label{eq:generalCollAvoid}
			d(\tilde{\mathbf{x}}_i^g(t),\tilde{\mathbf{x}}_\zeta^g(t))
			\geq
			\lambda\tilde{v}_i(t) + \underline{d} + \delta(t),
		\end{equation}
		where $\lambda$ is the so called
		\textit{time-headway} (measured in seconds)
		and
		$\underline{d}$ is the fixed
		\textit{bumper-to-bumper}
		distance defined in Section \ref{sec:vehModel}.
		The remaining term,
		$\delta(t)$, 
		is a slack variable
		that will be weighted in the cost function
		and that is traditionally
		used in the formulation of
		soft constraints.
		This allows
		the optimal controller to increase,
		if possible,
		the safety distance.
		In Section \ref{sec:otherConstr},
		$\delta(t)$ will be constrained
		to a given set.		
		
		Clearly, having $d(\tilde{\mathbf{x}}_i^g(t),\tilde{\mathbf{x}}_\zeta^g(t))$ 
		formulated as in (\ref{eq:distanceVeh})
		makes the problem
		non-convex.
		In order to prevent
		this, in \cite{molinari2018automation},
		collision 
		avoidance constraints
		were reformulated.
		By \cite[Proposition 3]{molinari2018automation},
		constraints to 
		avoid collisions with frontal
		vehicles 
		can be rewritten as
		\begin{multline}
			\label{eq:convexFrontalConstr}
			\forall t\in [0,T_h],\ \forall \zeta\in\mathcal{F}_i^t,\\
			\mathcal{M}_i^{-1}(\mathcal{M}_\zeta(\tilde{p}_\zeta(t)))-
			\tilde{p}_i(t)\geq \lambda\tilde{v}_i(t) + \underline{d} + \delta(t).
		\end{multline}
		Clearly, (\ref{eq:convexFrontalConstr}) is convex.
		
		In \cite{molinari2018automation},
		left turns are disallowed, since
		allowing left turns
		is claimed to lower
		traffic efficiency.
		In this paper,
		we do allow left turns
		at intersections
		and analyze
		the effect of this
		in Section \ref{sec:leftTurns}.
		First, we introduce 
		a general convex collision avoidance
		constraint 
		for vehicle $i$
		and all higher priority
		vehicles
		crossing
		$\mathcal{P}_i$ 
		i.e. $\zeta\in\mathcal{L}_i^k\setminus\mathcal{F}_i^k$. 
		The following proposition
		extends the outcome of
		\cite[Proposition 4]{molinari2018automation}.
		\begin{proposition}
			Given a vehicle $\zeta\in\mathcal{L}_i^k\setminus\mathcal{F}_i^k$,
			the constraint, $\forall h\in G_i^k\cap G_\zeta^k$:
			\begin{multline}
				\label{eq:collAvoidConstrProof}
				\mathcal{M}_i^{-1}(h)-\tilde{p}_i(t)\geq \lambda\tilde{v}_i(t) + \underline{d} + \delta(t),\\				
				\forall t\in [0,T_h]:\ \zeta\not\in\mathcal{F}_i^t \land \mathcal{M}_\zeta^{-1}(h)-\tilde{p}_\zeta(t)\geq-\underline{d},
			\end{multline}
			guarantees that any collision between $i$ and $\zeta$ 
			throughout the prediction horizon
			is avoided.
			
			\begin{proof}
				A collision between $i$ and $\zeta$ occurs at prediction time $t$ if, as in Section \ref{sec:vehModel},
				\begin{equation}
					\label{eq:defColl}
					d(\mathbf{x}_i^g(t),\ \mathbf{x}_\zeta^g(t))\leq\underline{d}.
				\end{equation}
				We can distinguish three different cases.
				(i) In the case $\zeta\in\mathcal{F}_i^t$,
				any collision is avoided if (\ref{eq:convexFrontalConstr}) holds.
				(ii) if $\zeta\not\in\mathcal{F}_i^t$,
				but $\mathcal{M}_\zeta^{-1}(h)-\tilde{p}_\zeta(t)<-\underline{d}$,
				vehicle $\zeta$ at prediction time $t$ 
				does not lie
				on $\mathcal{P}_i$
				and has driven 
				at a distance larger than $\underline{d}$
				far from $\mathcal{P}_i$,
				(\ref{eq:defColl}) does not hold.
				(iii) 
				as long as
				$\zeta\not\in\mathcal{F}_i^t \land \mathcal{M}_\zeta^{-1}(h)-\tilde{p}_\zeta(t)\geq-\underline{d}$,
				the distance between $i$
				and $\mathcal{P}_\zeta$ 
				is larger than $\lambda\tilde{v}_i(t) + \underline{d} + \delta(t)$,
				which is clearly larger than $\underline{d}$.
				This concludes the proof.
				\qed
			\end{proof}		
		\end{proposition}
		The resulting convex collision avoidance constraint then becomes:	
		\begin{multline}
			\label{eq:convexCrossConstr}
			\forall h\in G_i^k\cap G_\zeta^k,\ 
			\mathcal{M}_i^{-1}(h)-\tilde{p}_i(t)\geq \lambda\tilde{v}_i(t) + \underline{d} + \delta(t),\\				
			\forall t\in [0,T_h]:\ \zeta\not\in\mathcal{F}_i^t \land \mathcal{M}_\zeta^{-1}(h)-\tilde{p}_\zeta(t)\geq-\underline{d}.
		\end{multline}	
	\subsection{Other Constraints and Cost Function}
		\label{sec:otherConstr}
		Traditional
		constraints on the
		allowed speed and acceleration
		can be formulated as follows:
		\begin{align}
			\label{eq:classicConstr1}
			\forall t\in[0,T_h],\ &\tilde{u}_i(t)\in[\underline{a}_i,\bar{a}_i],
			\\
			\label{eq:classicConstr2}
			&\tilde{v}_i(t)\in[\underline{v},\bar{v}],
			\\
			\label{eq:classicConstr3}
			&\tilde{v}_\zeta(t)\in[\underline{v},\bar{v}],
		\end{align}
		where $\underline{v}>0$, 
		thus guaranteeing that
		vehicles do not drive backwards.
		
		The slack variable $\delta(t)$
		is also constrained to a given set, 
		i.e.
		\begin{equation}
			\label{eq:constrainedDelta}
			\forall t\in[0,T_h],\ 
			\delta(t)\in[-\bar{\lambda}\tilde{v}_i(t),\bar{\delta}],
		\end{equation}
		where $0\leq\bar{\lambda}<\lambda$ and $\bar{\delta}>0$.
		By this,
		the time headway in (\ref{eq:generalCollAvoid})
		can be diminished down to $\lambda-\bar{\lambda}$,
		although this will be negatively weighted in the cost function,
		thus aiming to increase the safety distance.
		
		The controller performance
		can be evaluated
		by the following cost function
		\begin{equation}
			\label{eq:costFcn}
			J = \sum_{t=0}^{T_h}
			q(\tilde{v}_i(t)-v_i^r)^2 +
			r\tilde{u}_i^2(t) + 
			\omega \delta(t),
		\end{equation}
		where $q,r\in\mathbb{R}_{>0}$ 
		and $\omega\in\mathbb{R}_{<0}$
		are design parameters.
		The first term of the sum
		punishes deviations from the desired speed $v_i^r$;
		the second term
		discourages high values
		of acceleration or deceleration,
		thus pledging a better comfort on-board.
		The last term punishes small $\delta(t)$,
		hence contributing to crease safety.
		
		The optimal control problem
		is then formulated as follows:
		\begin{align}
				\min\limits_{\tilde{u}_i(0)\dots\tilde{u}_i(T_h)}\ \ \  &\text{cost function (\ref{eq:costFcn})}\nonumber\\
				\text{s.t.}\ \ \   &\text{Prediction model (\ref{eq:predModel_i}-\ref{eq:initialCond2})}\nonumber\\
				&\text{Safety constraints (\ref{eq:convexFrontalConstr}, \ref{eq:convexCrossConstr})}\nonumber\\
				&\text{Input and state constraints (\ref{eq:classicConstr1}- \ref{eq:constrainedDelta})}\nonumber.
		\end{align}
		A vector of inputs composed of $\tilde{u}_i(0)\dots\tilde{u}_i(T_h)$
		will solve the optimal control problem.
		Coherently with the MPC structure,
		only the first optimal control input
		will be executed in (\ref{eq:pointMassLinSys})
		at each $k\in\mathbb{N}$, i.e.
		$u_i(k)=\tilde{u}_i(0)$.
\section{Simulation}
\label{sec:simulations}
	\begin{table}[t]	
		\vspace{10px}	 
		\caption{Problem data. }
		\label{tab:problemData}
		\begin{tabular}{c}
			\hline\\
			$L_w=3.5m$, $D_w=30m$, $T_s=0.25 s$, $p_c=1$, $p_d=0.1$, $\epsilon=0.1$\\
			$\lambda=1s$, $\bar{\lambda}=0.5s$, $\bar{\delta}=10m$, $\underline{d}=2.1m$, $\underline{v}=0$,  $\bar{v}=130km/h$,\\
			$T_h=10$, $\underline{a}_i=-9m/s^2$, $\bar{a}_i=5m/s^2$, $q=0.1$, $r=0.01$, $\omega=-0.1$\\
			$v_{min}=52km/h$, $v_{max}=56km/h$, $\wp=0.5$
			\\
			\hline
			\\
		\end{tabular}
	\end{table}
	\begin{figure}[t]
		\vspace{10px}
		\includegraphics[width=.85\columnwidth]{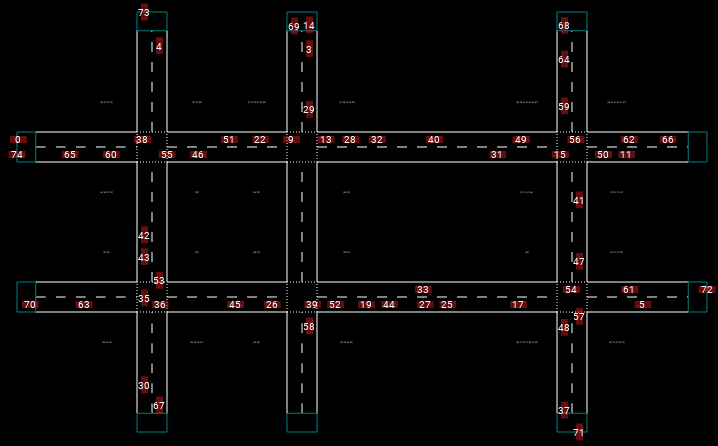}
		\centering
		\caption{Using the open-source software YatSim for modeling the urban road network.}
		\label{fig:yatsim}
	\end{figure}
	\begin{figure}[t]
		\begin{subfigure}[t]{\columnwidth}
			\includegraphics[width=\columnwidth]{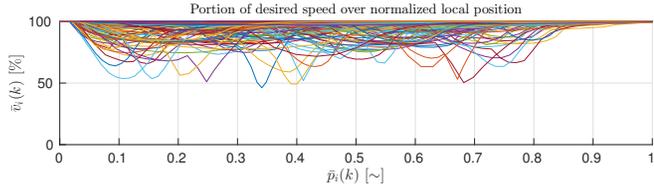}
			\caption{Vehicles' speed ratios over normalized local coordinate.}
			\label{fig:L_normSpeed_normPos}
		\end{subfigure}
		\begin{subfigure}[t]{\columnwidth}
			\includegraphics[width=\columnwidth]{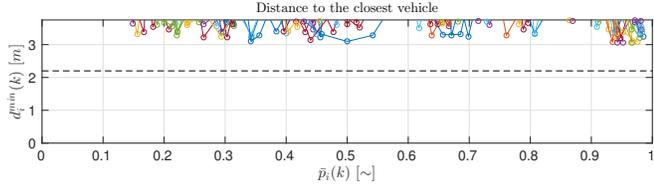}
			\caption{Vehicles' minimum distance over normalized local coordinate.}
			\label{fig:L_Dist2Closest}
		\end{subfigure}
		\begin{subfigure}[t]{\columnwidth}
			\includegraphics[width=\columnwidth]{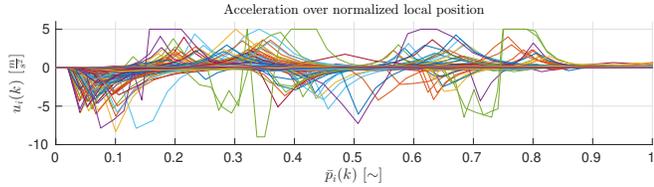}
			\caption{Vehicles' accelerations over normalized local coordinate. The dashed line indicates the minimum allowed distance,}
			\label{fig:L_accel_normPos}
		\end{subfigure}
		\caption{Analysis of individual vehicles' characteristics.}
		\label{fig:randomized_experiment}
	\end{figure}
	\begin{figure}[t]
		\begin{subfigure}[t]{\columnwidth}
			\includegraphics[width=\columnwidth]{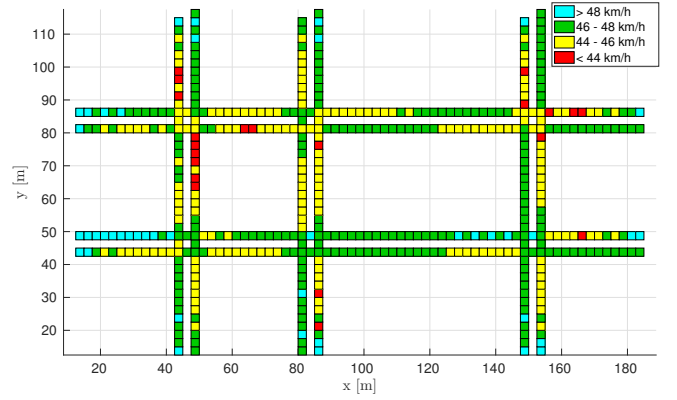}
			\caption{Macroscopic analysis of the average speed in road network sections.}
			\label{fig:L_2D_Speed}
		\end{subfigure}
		\begin{subfigure}[t]{\columnwidth}
			\includegraphics[width=\columnwidth]{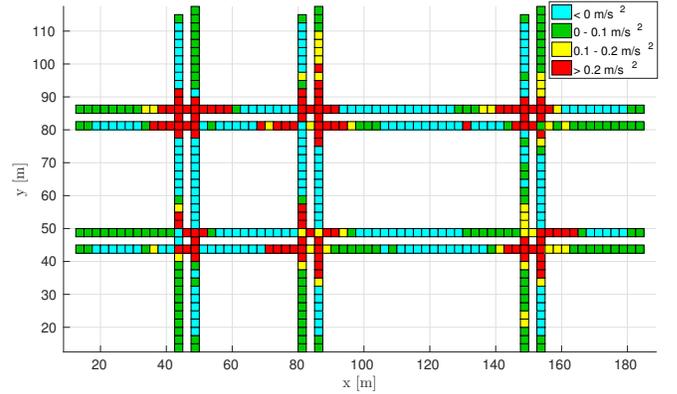}
			\caption{Macroscopic analysis of the average acceleration in road network sections.}
			\label{fig:L_2D_Accel}
		\end{subfigure}
		\caption{Macroscopic analysis.}
	\end{figure}
	The traffic flowing in the
	urban road network
	sketched in
	Figure \ref{fig:roadNet}
	can be simulated
	by means of the open-source programme
	YatSim \cite{dethof2019yatsim},
	as in Figure \ref{fig:yatsim}.
	Vehicles will employ the
	decentralized control strategy
	presented in this paper.
	In the following, 
	a macroscopic
	analysis
	of the traffic (in the sense of \cite{may1990traffic})
	will be conducted.
	Accordingly,
	the traffic will be studied as a flow.
	This allows to quantify
	the impact that this control strategy
	has on the urban road network as a whole,
	hopefully resulting in 
	an improvement of flow performance indexes,
	such as average or minimum speed and acceleration.	
	
	\subsection{Macroscopic Analysis}
	Vehicles
	will be randomly injected into the network;
	at each discrete time step from each road,
	provided that there is enough space to avoid trivial initial collisions,
	a vehicle, say $i\in\mathbf{N}$, will enter the road network
	with a given probability $\wp\in(0,1)$.
	Its desired speed $v_i^r$ 
	will be drawn out of an 
	uniform distribution 
	$\mathcal{U}_{[v_{min},v_{max}]}$.
	Also
	its desired path $\mathcal{P}_i$
	will be chosen in a random way,
	as illustrated 
	in \cite{dethof2019yatsim}.
	The parameters
	used for the simulation
	are contained in 
	Table \ref{tab:problemData}.
	The simulation is run until more than 
	$500$ vehicles complete their paths.
	For what concerning simulation results, 
	the average speed 
	of the traffic is 
	$46.8 [\frac{km}{h}]$,
	which,
	by \cite{cityspeed},
	is much higher
	than the current average traffic speed
	in New York City
	($28.32 [\frac{km}{h}]$)
	or Boston
	($33.64 [\frac{km}{h}]$).
	We use these two cities as benchmark,
	since the road network structure
	composed of perpendicularly intersecting
	streets is similar to the one considered here.
	Moreover, $\wp$ has been chosen
	to replicate real congestion conditions.
	Vehicles are injected in the
	simulation environment with their desired speeds.
	Intuitively, they will have to slow down
	in order to avoid collisions.
	In fact, the average acceleration
	of the traffic is
	$-0.0212 [\frac{m}{s^2}]$,
	thus showing a breaking behavior.
	
	Given a vehicle $i\in\mathbf{N}$,
	let $\bar{p}_i(k)$
	be its normalized
	local coordinate, i.e. $\forall k\in\mathbb{N}$,
	$$
	\bar{p}_i(k)
	=
	\frac{p_i(k)}{\max\limits_k p_i(k)}	
	\in
	[0,1]
	,
	$$
	let $\bar{v}_i(k)$
	be its speed ratio, i.e. $\forall k\in\mathbb{N}$,
	$$
	\bar{v}_i(k)
	=
	\frac{v_i(k)}{v_i^r}100\%	
	,
	$$
	and let $d_i^{min}(k)$ be the 
	minimum distance towards other vehicles
	at instant $k$, 
	i.e. $\forall k\in\mathbb{N}$,
	$$
	{d}_i^{min}(k)
	=
	\min\limits_{j\in \mathbb{N}\setminus \{i\} } d_{ij}(k)
	.
	$$
	Although this section presents 
	a macroscopic analysis,
	a detailed evaluation of individual vehicles'
	characteristics
	is also reviewed.
	Figure \ref{fig:L_normSpeed_normPos}
	shows the evolution
	of speed ratios
	as function of
	normalized local coordinates
	for each individual vehicle.
	The large majority of vehicles
	keep a speed 
	above the $80\%$ of
	their desired ones, 
	while
	no vehicle is
	slowing down below 
	the $48\%$ of its desired speed.
	By comparing this result to the 
	traditional solution with traffic lights,
	the improvement is explicit
	since no vehicle is actually stopping.
	This might intuitively result in 
	decreasing the traffic safety
	(which can be defined as the likelihood of having collisions).
	However, Figure \ref{fig:L_Dist2Closest}
	shows that all vehicles keep
	a safety distance larger than $\underline{d}$.
	Moreover, the amount of vehicles getting close
	to this minimum allowed value 
	is marginal.
	Figure \ref{fig:L_accel_normPos}
	outlines
	individual vehicles' accelerations
	as function of $\bar{p}_i(k)$.
	As already motivated above,
	many vehicles brake
	as soon as they enter
	the simulation environment.
	Vehicles are often saturating their acceleration, 
	in those cases when 
	collision avoidance constraints
	allow to do so 
	(e.g. as soon as that vehicle wins the decentralized auction).
	This saturating behavior can be intuitively mitigated
	by increasing $r$.
	Only in one point (e.g., for $p_i(k)$ between $0.3$ and $0.4$),
	a vehicle is saturating the braking power.
	This could be prevented by 
	increasing $T_h$,
	length of the prediction horizon, or
	$\omega$, weight of the hold safety distance.
	
	An analysis of which portions of the
	urban road network
	suffer the most of
	speed drops or
	robust braking
	is presented.
	First, 
	the road network
	has to be divided
	into small fragments 
	(e.g. squares of side $2.5[m]$)
	on which we compute the average
	characteristics of the traffic.
	As in Figure \ref{fig:L_2D_Speed},
	average speeds lower than $44[\frac{km}{h}]$
	are met in some roads entering an intersection 
	(e.g. $(x=50,y=77.5)$ or $(x=155,y=77.5)$),
	while, along those roads
	leaving an intersection,
	higher speeds are found.
	Figure \ref{fig:L_2D_Accel}
	shows that traffic along roads entering an intersection
	exhibits a braking behavior.
	On the other hand, vehicles try to leave the intersection
	as soon as possible once inside of it,
	thus resulting in higher average acceleration.
	\subsection{Impact of Left-Turns}
	\begin{figure}[t]
		\includegraphics[width=\columnwidth]{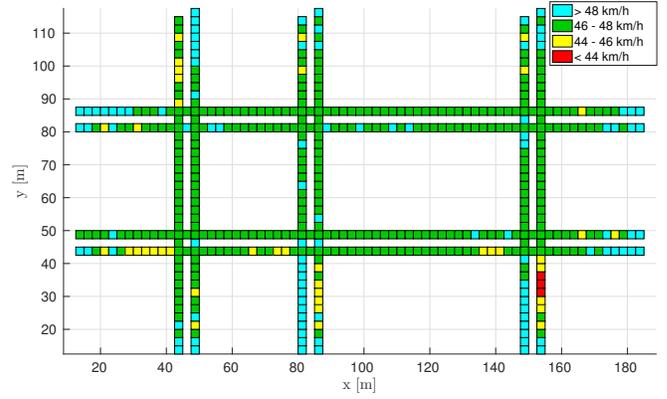}
		\caption{Macroscopic analysis of the average speed in road network sections. Left-turnings disallowed.}
		\label{fig:R_2D_Speed}
	\end{figure}			
	\begin{figure}[t]
		\includegraphics[width=\columnwidth]{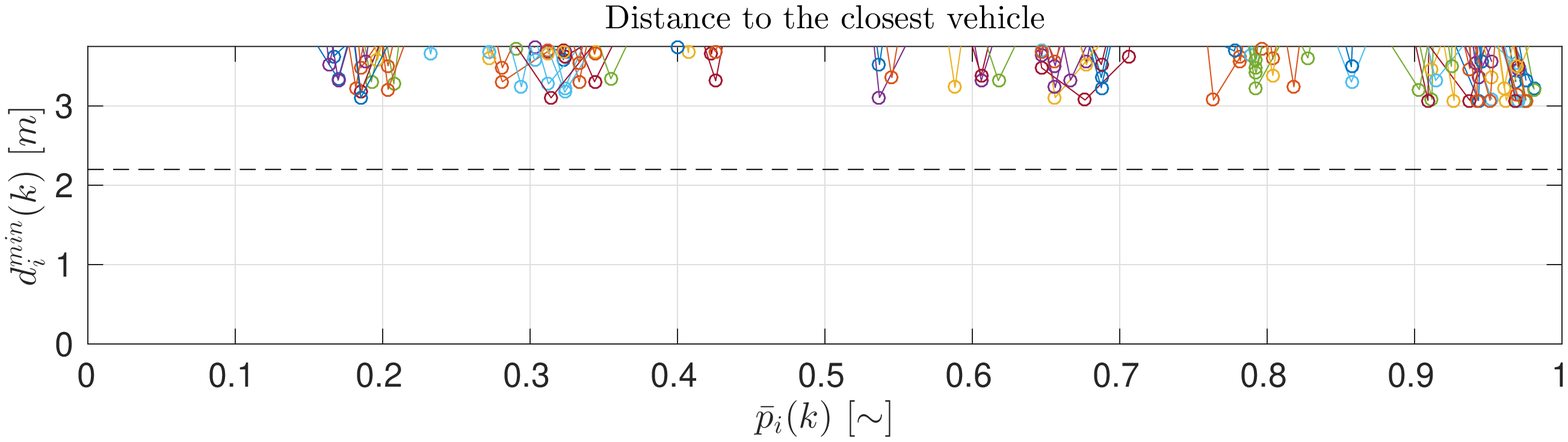}
		\caption{Vehicles' minimum distance over normalized local coordinate. Left turnings disallowed. The dashed line indicates the minimum allowed distance,}
		\label{fig:R_Dist2Closest}
	\end{figure}
	In \cite{molinari2018automation}, 
	left turnings were forbidden.
	Also, \cite{reich2012transportation} revealed that
	a world-wide known delivery
	company has historically prohibited left-turnings to
	its drivers.
	This was
	due to the assumption that allowing left-turns
	affects traffic efficiency
	in a negative way
	and increases hazards.
	In the following, we 
	repeat our simulation
	for the case when
	left-turns are forbidden.
	We simulate
	the same urban road network
	where cars have the same parameters
	(same initial and goal position)
	and are generated according to the same probability $\wp$;
	however,
	left turnings are 
	in this case
	not allowed.
	The average speed of the traffic
	is $47.8[\frac{km}{h}]$
	and the average acceleration
	is $-0.002[\frac{m}{s^2}]$.
	Additionally, Figure \ref{fig:R_2D_Speed}
	shows that in almost all the portions
	composing the road network,
	vehicles keep a higher speed than in
	Figure \ref{fig:L_2D_Speed}.
	With regards to the outcome of the previous section,
	it is therefore clear that 
	preventing vehicles from turning left
	increases the average speed
	and decreases their absolute acceleration.
	If the average speed is a measure
	of traffic efficiency,
	then the assumption presented by \cite{reich2012transportation}
	seems reasonable.
	
	On the other hand,
	under a safety-related point of view,
	we do not get to the same conclusion.
	Let us pick, as safety index,
	the minimum distance that each 
	vehicle keeps towards the others, and let this be plotted,
	as function of the normalized local coordinates,
	in Figure \ref{fig:R_Dist2Closest}.
	Its outcome is not
	much different
	than what
	Figure \ref{fig:L_Dist2Closest} yields.
	Therefore, left-turnings do not seem to 
	increase hazards in
	the autonomous road traffic.
	One can guess that,
	although the overall average speed is affected
	by the presence of a left-turning traffic,
	safety is anyhow guaranteed by the 
	presence of the on-board MPC controller.
	\label{sec:leftTurns}

\section{Conclusion}
	In this paper,
	we suggested the use of a decentralized
	control scheme for vehicles
	in a urban traffic network.
	This scheme is based on 
	a consensus-based protocol (CBAA-M),
	which gives vehicles crossing orders at the intersections,
	with an on-board MPC controller,
	responsible for avoiding collisions and tracking some performance.
	CBAA-M has been here proven to achieve
	an agreement in  cases
	when the network topology is connected.
	This is in contrast to \cite{molinari2018automation},
	where full connectedness was required.
	Finally, the impact of disallowing left-turning
	has been analyzed.
	
	Future work will focus
	on real-time implementability issues
	by improving the convergence rate
	of the consensus protocol
	and decrease the complexity of the on-board MPC.
\bibliography{bibliography}
\bibliographystyle{IEEEtran}

\end{document}